\documentclass[twocolumn,superscriptaddress,pra,aps,showpacs]{revtex4-1}

\usepackage{amsmath,amssymb}
\usepackage{graphicx}
\usepackage{bm}

\usepackage{color}
\usepackage{braket}

\usepackage{hyperref}
\hypersetup{
    colorlinks=true,       
    linkcolor=blue,        
    citecolor=blue,        
    filecolor=blue,        
    urlcolor=blue,         
    runcolor=blue
}

\newcommand{\notablue}[1]{\textcolor{black}{#1}}

\begin{document}

\title{Engineering strong coupling with molecular coatings in optical nanocavities}

\author{Athul S. Rema}

\author{Adri\'an E. Rubio López}

\affiliation{Department of Physics, Universidad de Santiago de Chile, Av. Victor Jara 3493, Santiago,  9170124, Chile.}

\author{Felipe Herrera}
\affiliation{Department of Physics, Universidad de Santiago de Chile, Av. Victor Jara 3493, Santiago,  9170124, Chile.}
\affiliation{Millennium Institute for Research in Optics, Concepci\'on, Chile.}
\email{felipe.herrera.u@usach.cl}

\date{\today}

\begin{abstract}
Quantum emitters near the surface of  silver nanoparticles undergo Rabi oscillations in electronic population dynamics due to strong coupling with near-field multipole modes that are not radiative. Low-frequency nanoparticle dipole modes are radiative but do not couple strong enough to quantum emitters. These features limit the observation of strong coupling.  Using macroscopic quantum electrodynamics theory within a Lorentzian pseudo-mode approximation for the non-Markovian interaction kernel, we demonstrate that by coating spherical silver nanoparticles with a thin molecular J-aggregate layer, the resulting core-shell plexciton resonance restructures the local electromagnetic vacuum at dipole-mode frequencies to enable Rabi oscillations for  quantum emitters that otherwise would only undergo exponential population decay. Specifically, we show for quantum dot emitters in the near field of silver nanospheres of 20 nm radius, that weak-to-strong coupling crossovers can be induced using 2 nm J-aggregate shells. Our work demonstrates the potential of molecular aggregates to enable deep sub-wavelength structuring of the vacuum field for the observation of coherent quantum dynamics in optical nanocavities.
\end{abstract}

\maketitle

\section{\label{sec:intro}Introduction}

Achieving strong coupling between a quantum emitter and a confined electromagnetic mode remains a central milestone in nanophotonics \cite{Torma2014,GarciaVidal2021}. In this regime, coherent energy exchange overcomes the intrinsic decay pathways of the cavity and emitter, enabling control over light-matter interactions at the nanoscale~\cite{Baranov2018,GonzalezTudela2024}. Plasmonic nanocavities are natural platforms for reaching strong coupling because they concentrate optical fields into extremely small volumes \cite{Haran2019,Park2019,Chikkaraddy2016,Santhosh2016}. Placing emitters in close proximity to metal structures inside sub-nanometer gaps can in principle affect the emitter orbital structure, requiring treatments that go beyond simple point-dipole approximations~\cite{Neuman2018,Fregoni2021}. Extreme field confinements can also be detrimental to quantum coherence, due to non-radiative population quenching into the metal structure~\cite{Anger2006,Hohenester2008,Delga2014,Pelton2015,Khurgin2015}. Therefore, strategies to reduce emitter-metal distances without losing quantum coherence are important. 

An alternative to modifying the metal-emitter distance for reaching strong coupling is to restructure the photonic environment of optical nanocavities. In this work, we propose  coating a metallic nanoparticle with a highly polarizable molecular aggregate layer to restructure the local density of optical states experienced by a nearby quantum emitter. Molecular coatings can produce plexcitonic resonances due to strong interaction between molecular excitons and surface plasmons~\cite{Fofang2008,Fofang2011,Wersall2017,Wiederrecht2004,Rider2022}. The optical response of the underlying nanoparticle is substantially altered if the coating material possesses a large collective oscillator strength. Strong optical absorption and the resulting anomalous dispersion in the coating can shift the boundary conditions of the local electromagnetic field, forming new resonant plexcitonic modes that are not naturally present in bare metallic structures \cite{Antosiewicz2014}.

Modeling the non-Markovian dynamics of quantum emitters in the near field of highly-structured plexcitonic resonances requires  macroscopic quantum electrodynamics theory \cite{Buhmann2012,Feist2021}, which can consistently describe weak coupling phenomena (Purcell-enhanced emission) and strong coupling physics (light-matter hybridization), for realistic nanocavity structures composed of dispersive and absorptive materials. This theoretical framework was used recently to study the steady-state photon statistics of plexcitons in specialized mirror-gap geometries~\cite{SaezBlazquez2022}. Here, we focus on the transient dynamics of the emitter electronic population induced by core-shell spherical nanoparticles. Standard Markovian quantum optics theory fails to fully capture the coupled emitter-nanoparticle system dynamics, due to the limitations of Markovian theory for treating broadband multi-mode nanocavity structures \cite{Herrera2020,Herrera2024}.

We apply a recently developed Lorentzian kernel approximation \cite{Rema2025} for solving the non-Markovian quantum dynamics equations that govern the emitter-nanoparticle system. This approximation becomes exact for nanocavities composed of Drude metal nanoparticles, which produce simple Lorentzian resonances in the local density of photon states \cite{Medina2021}, but can accurately solve the quantum dynamics of emitters in more complex systems that involve strong photon-photon interactions between near-field modes \cite{Rema2025}. Using this method, we  demonstrate that by inserting an active molecular J-aggregate layer in the gap between the surface of a silver metal sphere and a quantum dot emitter, the electromagnetic vacuum at the emitter position is modified such that over frequency regions where Purcell-enhanced spontaneous emission is expected without the aggregate layer, Rabi oscillations of the population due to strong coupling occur, associated with large Rabi splittings in the emission spectrum. Our work provides the first rigorous demonstration that molecular coatings can be used for engineering the quantum  dynamics of dipole emitters at target frequencies, without modifying the distance between the emitter and the metallic surface.

The rest of the article is organized as follows: Section~\ref{sec:theory} introduces the theoretical framework, detailing the macroscopic quantum electrodynamics formalism and the Lorentzian kernel approximation. Section~\ref{sec:system} defines the plexcitonic nanoparticle system, comparing the material permittivities and hybridization mechanisms of the bare and coated silver nanospheres. Section~\ref{sec:results} presents our main results, first evaluating the restructured spectral density of the electromagnetic vacuum, and then analyzing the resulting time-domain population dynamics and coherence spectra. Finally, Section~\ref{sec:conclusions} summarizes our findings and provides an outlook on experimental feasibility and future theoretical extensions.

\section{\label{sec:theory}Theoretical framework}

\subsection{\label{sec:macroqed}Macroscopic QED and the memory kernel}

We treat light--matter interaction using macroscopic QED theory~\cite{Buhmann2012,Scheel2008}. The dipole emitter is described as a charge distribution and the cavity is defined by the confined near field of an arbitrary material structure with local linear permittivity $\epsilon(\omega)$ and permeability $\mu(\omega)$. The total Hamiltonian derived from first principles can be written as~\cite{Feist2021,Wang2019,Medina2021,Schafer2024, Rema2025}
\begin{equation}\label{eq:total H}
  \hat{\mathcal{H}} = \hat{\mathcal{H}}_\mathrm{dip}
  + \hat{\mathcal{H}}_\mathrm{f}
  + \hat{\mathcal{H}}_\mathrm{int}\,,
\end{equation}
with the qubit described by $\hat{\mathcal{H}}_\mathrm{dip} = \hbar\omega_g\lvert g\rangle\langle g\rvert + \hbar\omega_e\lvert e\rangle\langle e\rvert$, where $\lvert g\rangle$ and $\lvert e\rangle$ are the ground and excited dipole states, with frequencies $\omega_g$ and $\omega_e$, respectively. The quantized near field is described by
\begin{equation}\label{eq:field H}
  \hat{\mathcal{H}}_\mathrm{f} = \int_0^\infty d\omega\,
  \hbar\omega\,\hat{a}^\dagger(\omega)\hat{a}(\omega)\,,
\end{equation}
with field operators satisfying bosonic commutation relations $[\hat{a}(\omega),\hat{a}^\dagger(\omega')] = \delta(\omega-\omega')$ and $[\hat{a}(\omega),\hat{a}(\omega')] = 0$. These are defined from the fundamental field modes in macroscopic QED through an emitter-centered orthonormalization procedure introduced in Ref.~\cite{Feist2021}.

The light--matter interaction Hamiltonian is given by
\begin{equation}\label{eq:light-matter H}
  \hat{\mathcal{H}}_\mathrm{int} = \int_0^\infty d\omega\,
  g(\omega)\,\hat{d}^{(+)}\hat{a}(\omega) + \text{H.c.}\,,
\end{equation}
with the coupling function given by
\begin{equation}\label{eq:g coupling}
  g(\omega) = \sqrt{\frac{\hbar\omega^2}{\pi\epsilon_0 c^2}\,
  \mathbf{e}_\alpha \cdot
  {\rm Im}[\stackrel{\leftrightarrow}{\mathbf{G}}
  (\mathbf{r}_0,\mathbf{r}_0,\omega)]
  \cdot \mathbf{e}_\alpha}\,,
\end{equation}
where $\stackrel{\leftrightarrow}{\mathbf{G}}(\mathbf{r}_0,\mathbf{r}_0,\omega)$ is the dyadic Green's tensor~\cite{Chew1999} evaluated at the emitter position $\mathbf{r}_0$ for dipole orientation $\mathbf{e}_\alpha$. The dipole operator is $\hat{d}^{(+)} = d_{eg}\lvert e\rangle\langle g\rvert$, with $d_{eg} = \langle e\rvert\hat{d}\rvert g\rangle$ the transition dipole moment. Equation~(\ref{eq:light-matter H}) is in the rotating-wave approximation. Operating within this approximation allows us to construct a semi-analytical solution for the non-Markovian dynamics, which is useful for modeling structured nanophotonic environments~\cite{Rema2025}.

We use the following Wigner--Weisskopf ansatz for the dipole-photon state $\lvert\psi(t)\rangle$,
\begin{eqnarray}\label{eq:wf ansatz}
  \lvert\psi(t)\rangle &=&
  C_{g0}(t)\lvert g\rangle\lvert\{0\}\rangle
  + C_{e0}(t)\,\mathrm{e}^{-i\omega_e t}
  \lvert e\rangle\lvert\{0\}\rangle
  \nonumber\\
  &&+ \int_0^\infty d\omega\,C_{g1}(\omega,t)\,
  \mathrm{e}^{-i\omega t}
  \lvert g\rangle\lvert\{\mathbf{1}(\omega)\}\rangle\,,
\end{eqnarray}
where $\lvert\{0\}\rangle$ is the photonic vacuum and $\lvert\{\mathbf{1}(\omega)\}\rangle = \hat{a}^\dagger(\omega)\ket{\{0\}}$ is the one-photon eigenstate of $\hat{\mathcal{H}}_\mathrm{f}$. $C_{e0}(t)$ is the dipole excited-state amplitude and $C_{g1}(\omega,t)$ is the 
single-photon wavefunction~\cite{Sipe1995,Fedorov2005}.

We consider spontaneous emission from an initially excited emitter with no photons present. In the absence of external driving, the ground-state amplitude $C_{g0}(t)$ decouples from the remaining system dynamics~\cite{Rema2025} (see Appendix~\ref{app:ide}). The relevant initial conditions governing the evolution are therefore $C_{e0}(0)=1$ and $C_{g1}(\omega,0)=0$.

Projecting the Schr\"odinger equation onto the basis states yields an integro-differential equation (IDE) for the dipole excitation amplitude (see derivation in Appendix~\ref{app:ide})
\begin{equation}\label{eq:Ce0}
  \dot{C}_{e0}(t) = -\int_0^t dt'\,
  \mathcal{K}(t-t')\,C_{e0}(t')\,,
\end{equation}
with the evolution governed by the delay kernel
\begin{equation}\label{eq:delay kernel}
  \mathcal{K}(t-t') = \int_0^\infty d\omega\,
  \mathcal{K}(\omega)\,
  \mathrm{e}^{-i(\omega-\omega_e)(t-t')}\,.
\end{equation}
The kernel spectrum $\mathcal{K}(\omega) = d_{eg}^2|g(\omega)|^2/\hbar^2$ (units of Hz) is sometimes referred to as spectral density in the literature. It encodes the full broadband electromagnetic response at the emitter location and is computed exactly from the dyadic Green's function. The single-photon amplitude follows by direct integration,
\begin{equation}\label{eq:Cg1}
  C_{g1}(\omega,t) = -i\int_0^t dt'\,
  \sqrt{\mathcal{K}(\omega)}\,
  C_{e0}(t')\,\mathrm{e}^{i(\omega-\omega_e)t'}\,.
\end{equation}

\notablue{The expectation value of the quantized electric field $\hat{\mathbf{E}}(\mathbf{r}_0,t)$ at the location of the emitter can be measured in amplitude and phase using near-field microscopy techniques \cite{Raschke2023,Esses2026}. Assuming that the emitter dipole is aligned with $\hat{\mathbf{e}}_\alpha$ and the ground state remains undepleted ($C_{g0}(t)\approx 1$), the near-field in the Wigner-Weisskopf ansatz approximation can be written as \cite{Rema2025}
\begin{equation}\label{eq:E amp}
\langle  \hat{\mathbf{E}}(\mathbf{r}_0,t)\rangle=  \int_0^\infty d\omega\, \mathbf{E}(\mathbf{r}_0,\omega)\, C_{g1}(\omega,t)\mathrm{e}^{-i\omega t}
\end{equation}
with Fourier amplitude given by
\begin{equation}\label{eq E vector single}
\mathbf{E}(\mathbf{r}_0,\omega)=\frac{\hbar \omega^2}{\pi \epsilon_0 c^2} \frac{{\rm Im}[\stackrel{\leftrightarrow}{\mathbf{G}}(\mathbf{r}_0,\mathbf{r}_0,\omega)]}{g(\omega)}\cdot \hat{\mathbf{e}}_\alpha.
\end{equation} 
}

\notablue{From the excited state amplitude in the Schr\"odinger frame $\tilde C_{e0}(t)\equiv C_{e0}(t){\rm exp}[-i\omega_et ]$, the complex dipole coherence spectrum can be written as
\begin{equation}\label{eq:coherence spectrum}
S_{\rm d}(\omega)=\int_{-\infty}^\infty dt \,{\rm Im}\tilde{C}_{e0}(t){\rm e}^{-i\omega t}.
\end{equation}
This quantity describes the spectral content of the coupled dipole-nanoparticle coherence, including phase information. In this work, we refer to the magnitude $|S_{\rm d}(\omega)|$ as  ``coherence spectrum",  ignoring phase information. In Appendix \ref{app:ide}, we show that the magnitude is sufficient to quantify near-field Rabi splittings. Analysis of the coupled far-field spectrum would require quantum modeling of the decay of the broadband photon state, along the lines described in Ref. \cite{SaezBlazquez2022}, which is beyond the scope of this work.}

\subsection{\label{sec:lorentzian}Lorentzian kernel approximation}

Equation~\eqref{eq:Ce0} can be solved numerically using standard integration methods~\cite{LEATHERS2005}, once the delay kernel $\mathcal{K}(\tau)$ is known. Analytical solutions are possible when the kernel in the time domain is given by a sum of exponentials~\cite{Polyanin2008handbook,BERMAN2010}. For a single exponential memory kernel of the form $\mathcal{K}(\tau) = A\,\mathrm{e}^{-\tilde{B}\tau}$, where $A$ is real and $\tilde{B}$ is complex valued, the Laplace domain solution of Eq~\eqref{eq:Ce0}, $G(s)=(s+\tilde{B})/(s^2+\tilde{B}s+A)$ inverts to~\cite{Rema2025}
\begin{equation}\label{eq:Ce0 analytical}
  C_{e0}(t) = \mathrm{e}^{-\tilde{B}t/2}
  \left[\cos(bt)+\frac{\tilde{B}}{2b}\sin(bt)\right]\,,
\end{equation}
with $2b=\sqrt{4A-\tilde{B}^2}$. Qualitatively, for a kernel with large enough amplitude and a small memory decay rate ($|A| \gg |\tilde{B}|^2/4$), the parameter $b$ is real and the system exhibits damped Rabi oscillations, a signature of strong coupling, with population lifetime $T_1 = 1/\mathrm{Re}[\tilde{B}]$ and Rabi frequency $\Omega_R = 2\mathrm{Re}[b]$. If the kernel amplitude is very small ($|A| \ll |\tilde{B}|^2/4$), then $C_{e0}$ undergoes simple exponential decay from its initial amplitude, a signature of weak coupling. 

The generalization of this analytical \notablue{single-mode} solution can be worked out by writing the memory kernel as the parametric model
\begin{equation}\label{eq:memory multimode}
  \mathcal{K}(\tau) = \sum_{j=1}^n A_j\,
  \mathrm{e}^{(-B_j-i[\Omega_j-\omega_e])\tau}\,,
\end{equation}
whose Fourier transform gives the Lorentzian expansion
\begin{equation}\label{eq:kernel lorentzian}
  \mathcal{K}(\omega) = \sum_{j=1}^n
  \frac{A_j}{\pi}\,
  \frac{B_j}{(\omega-\Omega_j)^2+B_j^2}\,,
\end{equation}
where each Lorentzian is centered at $\Omega_j$ with bandwidth $2B_j$ (FWHM) and area $A_j$. This expansion is used to approximate the true kernel spectrum $\mathcal{K}(\omega)$ through a fitting procedure, with $(n,A_j,B_j,\Omega_j)$ as fit parameters. When the near-field multipole expansion is converged, the exact memory kernel for the systems studied here yields exclusively positive amplitudes ($A_j > 0$). \notablue{In this case, Eq. (\ref{eq:memory multimode}) could be regarded as a physical pseudo-mode expansion~\cite{Garraway1997, Dalton2001}. In the most general case, the Lorentzian expansion has no specific physical interpretation other than a convenient representation that facilitates solution of an IDE. Following Ref. \cite{Rema2025}, the criterion $2A_j/B_j^2 > 1$ for strong coupling involving a particular pseudo-mode index $j$ could be considered. However, we discuss below that such mode-related metrics are not appropriate in multi-mode scenarios and offer an alternative analysis based on the Laplace solution for the wave function amplitudes.}

\begin{figure}[t]
  \includegraphics[width=0.8\columnwidth]{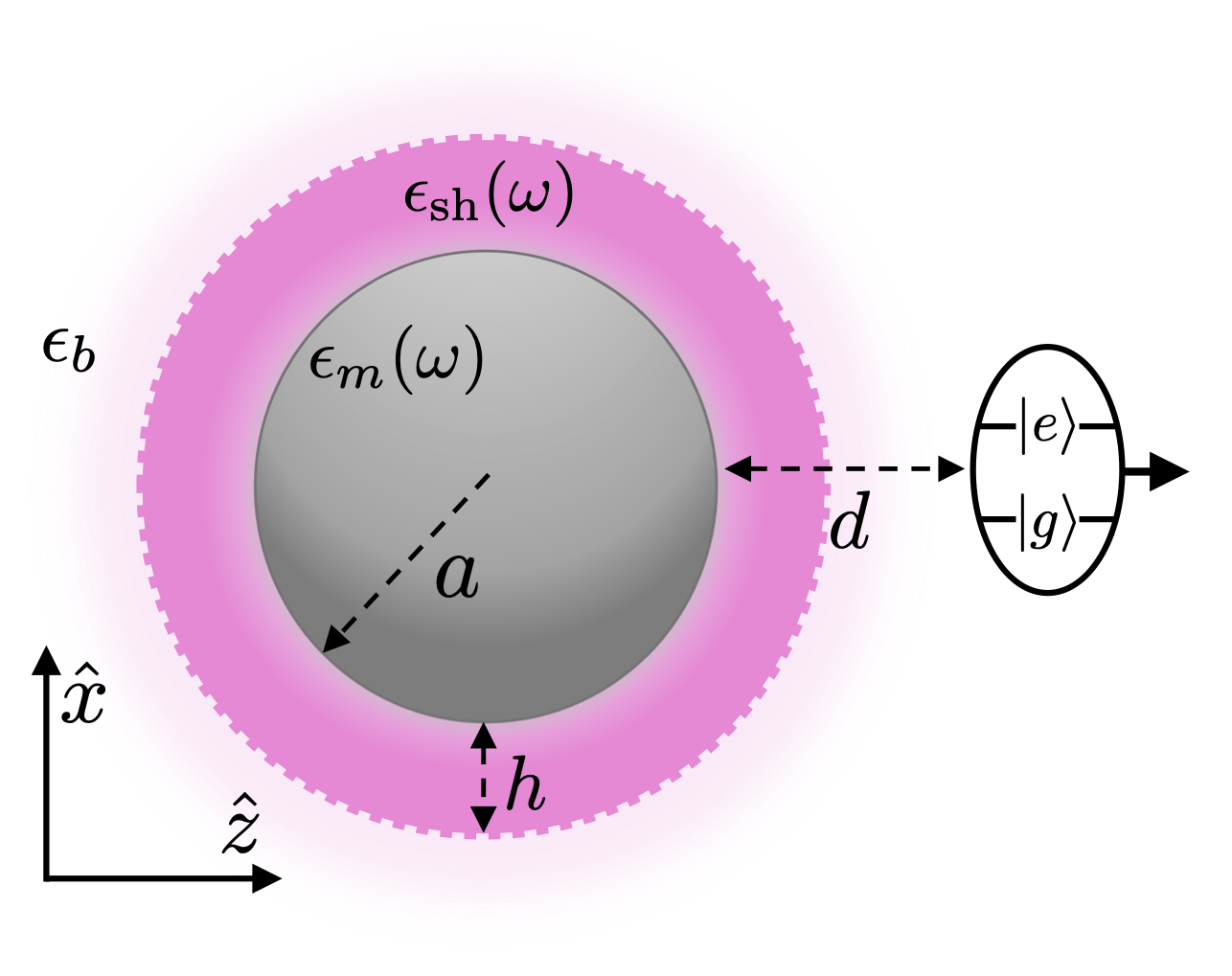}
  \caption{\label{fig:schematic}Schematic of the core--shell plexcitonic nanocavity. A silver nanosphere of radius $a=20$~nm is coated with a J-aggregate shell of thickness $h$ and suspended in a homogeneous medium ($n_b=1.3$, $\epsilon_b=1.69$). A $z$-oriented quantum dot ($d_{eg}=24$~D) is placed on the symmetry axis at a fixed metal-to-emitter gap of $d=3$~nm. The coated configuration features a shell thickness of $h=2$~nm ($\omega_\mathrm{ex}=\omega_\mathrm{sp}$, $f=0.3$, $\gamma_\mathrm{ex}=50$~meV).}
\end{figure}

\begin{figure*}[t]
  \includegraphics[width=\textwidth]{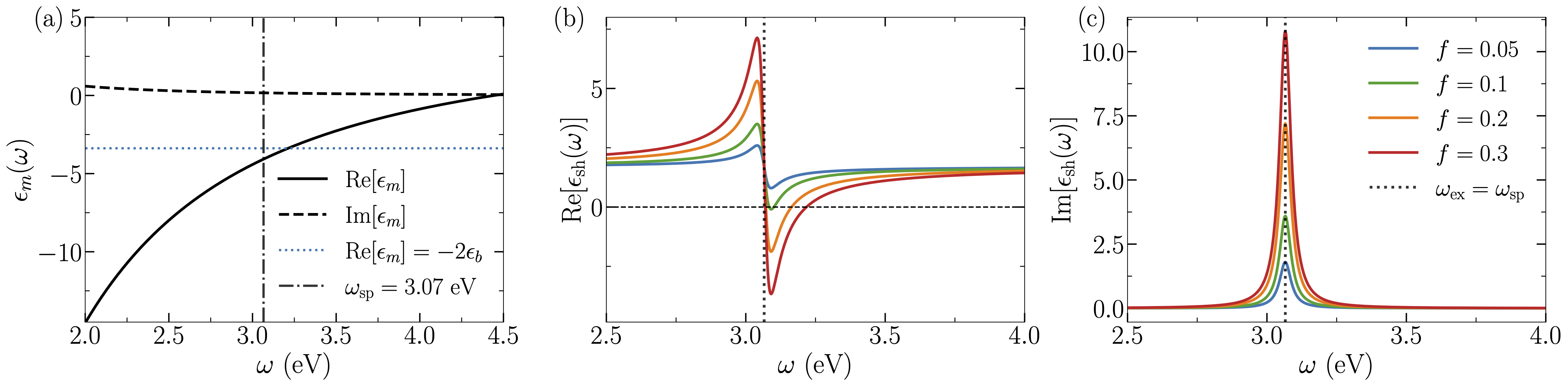}
  \caption{\label{fig:permittivity}Material permittivities. (a) Real (solid) and imaginary (dashed) parts of the Drude permittivity $\epsilon_m(\omega)$ of the silver core. The horizontal line marks the quasistatic Fr\"ohlich condition $\mathrm{Re}[\epsilon_m] = -2\epsilon_b$. The vertical dash-dotted line indicates the retarded dipole resonance $\omega_\mathrm{sp}\approx 3.07$~eV for the $a=20$~nm sphere, slightly red-shifted from the quasistatic intersection due to electrodynamic retardation. (b) Real and (c) imaginary parts of the J-aggregate shell permittivity $\epsilon_\mathrm{sh}(\omega)$ for varying nominal oscillator strengths $f$. In both (b) and (c), the exciton is tuned to the bare plasmon resonance ($\omega_\mathrm{ex} = \omega_\mathrm{sp}$) with linewidth $\gamma_\mathrm{ex} = 50$~meV, and the vertical dotted line marks $\omega_\mathrm{ex}$.}
\end{figure*}

Defining $\tilde{B}_j = B_j + i(\Omega_j-\omega_e)$, the Laplace transform of the IDE with $C_{e0}(0)=1$ yields
\begin{equation}\label{eq:laplace}
  \bar{C}_{e0}(s) = \frac{Q(s)}{P(s)}\,,
\end{equation}
where $Q(s) = \prod_k(s+\tilde{B}_k)$ and $P(s) = s\,Q(s)+\sum_k A_k\,Q_k(s)$ with $Q_k(s)=\prod_{j\neq k}(s+\tilde{B}_j)$. $P(s)$ is a polynomial of degree $n+1$. Its roots $\{s_m\}$ give the time-domain solution via the residue theorem~\cite{Polyanin2008handbook,Rema2025} (see Appendix~\ref{app:laplace solution}),
\begin{equation}\label{eq:residues}
C_{e0}(t) = \sum_m R_m\,\mathrm{e}^{s_m t}\,.
\end{equation}
Each pole $s_m = -\gamma_m + i\omega_m$ contributes a damped oscillation with residue weight $R_m = Q(s_m)/P'(s_m)$. The physical coupling regime is classified by the distribution of these residue weights. \notablue{We show below (Sec. \ref{sec:results}) that by analyzing the poles $s_m$ with most significant weight and their frequencies $\omega_m$ relative to the characteristic bandwidth of the kernel $\mathcal{K}(\omega)$, it is possible to establish criteria for single-mode and multi-mode strong coupling.}

\section{\label{sec:system}Plexcitonic nanoparticle system}

We study two nanocavity configurations for a single two-level emitter near a silver nanosphere. Figure~\ref{fig:schematic} illustrates the geometry. The system is suspended in a homogeneous  medium with refractive index $n_b=1.3$. The emitter has a transition dipole moment $d_{eg}=24$~D oriented along the $z$-axis. This dipole moment is representative of colloidal quantum dots~\cite{Savasta2010,Vanvlack2012}. The emitter is placed at a fixed distance $d=3$~nm from the metal surface. For the coated configuration, a 2~nm J-aggregate shell occupies the inner portion of the metal-emitter gap.

\begin{figure}[t]
  \includegraphics[width=0.5\textwidth]{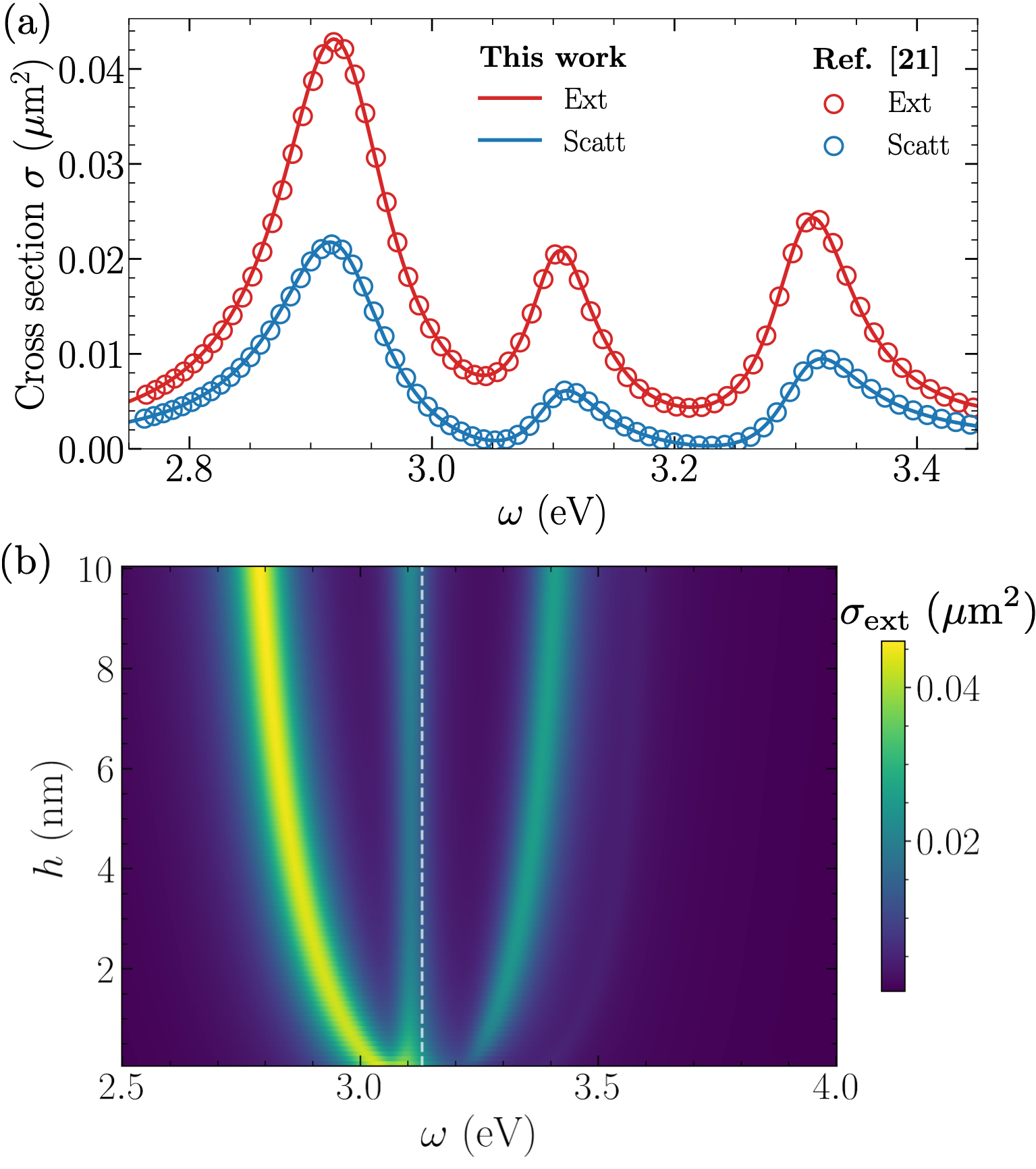}
  \caption{\notablue{ (a) Extinction and scattering cross sections of the Ag/J-aggregate core--shell nanoparticle (core radius $a = 20$~nm, shell thickness $h = 2$~nm). Solid curves represent our Lorenz--Mie calculations using the Drude core and Lorentz shell permittivities with the oscillator strength rescaling described in the text. Symbols denote results from Ref.~\cite{Antosiewicz2014}. The three peaks correspond, in order of increasing frequency, to the lower polariton, the geometric shell resonance, and the upper polariton; (b) Density plot with the extinction cross section for different values of the shell thicknesses $h$, for fixed core radius $a = 20$~nm.}}
  \label{fig:extinction}
\end{figure}

The silver core is described by the Drude permittivity
\begin{equation}\label{eq:eps drude}
  \epsilon_m(\omega) = \epsilon_\infty
  - \frac{\omega_p^2}{\omega(\omega+i\gamma)}\,,
\end{equation}
with $\epsilon_\infty=3.7$, plasma frequency $\omega_p=8.55$~eV, and damping rate $\gamma=65$~meV~\cite{Antosiewicz2014}. Mie theory places the dipolar localized surface plasmon resonance for a 20~nm sphere at $\omega_\mathrm{sp}\approx 3.07$~eV, slightly 
red-shifted from the quasi-static Fr\"ohlich condition ($\mathrm{Re}[\epsilon_m] = -2\epsilon_b$) due to finite-size retardation and radiation damping.

The second configuration coats the silver core with a thin excitonic shell modeled as a Lorentz oscillator, with permittivity
\begin{equation}\label{eq:eps shell}
  \epsilon_\mathrm{sh}(\omega)
  = \epsilon_{\infty,\mathrm{sh}}
  + \frac{f\,\omega_\mathrm{ex}^2}
  {\omega_\mathrm{ex}^2-\omega^2
  -i\gamma_\mathrm{ex}\omega}\,.
\end{equation}
Following Ref.~\cite{Antosiewicz2014}, we set the background permittivity $\epsilon_{\infty,\mathrm{sh}}=1.69$ matching the surrounding medium, shell thickness $h=2$~nm, exciton linewidth $\gamma_\mathrm{ex} = 50$~meV, and the exciton frequency $\omega_\mathrm{ex}$ tuned to the bare plasmon resonance $\omega_\mathrm{sp}$ to maximize hybridization. $f$ is the dimensionless oscillator strength.

A subtlety arises when tuning $\omega_\mathrm{ex}$ across different plasmon frequencies. At resonance, $\mathrm{Im}[\epsilon_\mathrm{sh}] = 
f\,\omega_\mathrm{ex}/\gamma_\mathrm{ex}$, which scales linearly with the exciton frequency. Simply shifting $\omega_\mathrm{ex}$ to track different plasmon modes would unphysically inflate the peak absorption. Following Ref.~\cite{Antosiewicz2014}, we correct this by defining $f$ at a fixed reference frequency $\omega^* = 1.8$~eV and rescaling the effective oscillator strength as $f \to f\,\omega^*/\omega_\mathrm{ex}$ in Eq.~(\ref{eq:eps shell}). The nominal value $f=0.3$ translates to an effective strength of $\tilde{f} \approx 0.18$ at the plasmon frequency, within reported experimental limits for cyanine dyes~\cite{Schlather2013}.

The critical material requirement for this coating is that $\mathrm{Re}[\epsilon_\mathrm{sh}]$ must drop below zero in a spectral window just above the exciton frequency. Any Lorentz oscillator achieves this if the oscillator strength is large enough and the linewidth is narrow. Prior quasistatic calculations established that $f \gtrsim 0.1$ with $\gamma_\mathrm{ex}=50$~meV is sufficient~\cite{Antosiewicz2014}. Molecular J-aggregates are well suited to meet this demand. Their superradiant collective transitions concentrate high oscillator strength into narrow 
spectral lines at room temperature~\cite{Schlather2013,Zengin2015}, routinely achieving $f \geq 0.3$ with exciton linewidths near $\gamma_\mathrm{ex} \approx 50$~meV due to motional narrowing. Cyanine dyes such as TDBC are frequently grown as conformal thin shells directly around colloidal nanoparticles~\cite{Fofang2008,Fofang2011}. Because the macroscopic oscillator strength scales with both the molecular packing density and the square of the molecular transition dipole~\cite{Rider2024}, dense coherently coupled layers naturally satisfy the conditions required to achieve $\mathrm{Re}[\epsilon_\mathrm{sh}]<0$ at visible frequencies. Transition metal dichalcogenide monolayers~\cite{Schneider2018} and perovskite quantum wells~\cite{Fieramosca2019} could also reach these conditions owing to their comparably large excitonic oscillator strengths.

Figure~\ref{fig:permittivity} displays the core and shell permittivities. When $f=0.3$, $\mathrm{Re}[\epsilon_\mathrm{sh}]$ drops below zero in a spectral window just above $\omega_\mathrm{ex}$. This transient metallic character alters the macroscopic optical response of the nanoparticle. As shown in the far-field cross sections in \notablue{Figure~\ref{fig:extinction}(a)}, the extinction and scattering spectra split into upper and lower polariton branches separated by $\sim 0.6$~eV, and a third narrow peak emerges directly between them. Our Lorenz--Mie calculations quantitatively agree with previous results for this geometry in Ref.~\cite{Antosiewicz2014}. 

\subsection{\label{sec:geometric}Geometric mode and the restructured vacuum}

To understand the origin of the \notablue{middle scattering peak in Fig.~\ref{fig:extinction}(a)}, we consider the plasmon hybridization framework~\cite{Prodan2003}. Any nanoscale shell supports two boundary resonances: an outer sphere mode and an inner cavity (or void) mode. These hybridize into bonding and antibonding states whose splitting depends on the shell's aspect ratio and the permittivity contrast across its interfaces~\cite{Gulen2013}. For a standard dielectric shell, these hybridized modes carry minimal oscillator strength and produce no sharp optical features. The situation changes when the molecular layer has negative real part of the permittivity just above $\omega_\mathrm{ex}$, which gives the shell an effectively metallic character in that frequency window. \notablue{This hybridizes the void mode into a sharp optical resonance, which is denoted here as  ``geometric mode", following the assignment in Ref.~\cite{Antosiewicz2014}.}

Several properties distinguish the geometric mode from the standard polariton branches. Its linewidth (FWHM $\approx 50$~meV) tracks the 
excitonic damping $\gamma_\mathrm{ex}$ rather than the broader polariton linewidths, which inherit additional contributions from radiation damping and the plasmon's multipole structure. A finite shell thickness gives it large spectral weight by concentrating surface charges almost exclusively at the outer interface. \notablue{Figure~\ref{fig:extinction}(b) shows that the spectral position of the geometric mode ($\sim 3.14$ eV) is independent of the shell thickness. The spectral position is also independent of the silver sphere radius.}

The far-field cross sections characterize the nanoparticle's antenna response. The near-field quantity that governs the emitter's quantum dynamics is the imaginary part of the Green's tensor, $\mathrm{Im}[G_{zz}(\mathbf{r}_0,\mathbf{r}_0,\omega)]$, evaluated at the emitter position. This narrow geometric mode creates a sharp peak in $\mathrm{Im}[G_{zz}]$ precisely where the bare metal provides only a broad, featureless spectral density. The kernel spectrum $\mathcal{K}(\omega)$ in Eq.~(\ref{eq:Ce0}) is constructed directly from this Green's tensor via Eq.~(\ref{eq:g coupling}). We compute it using analytic Mie theory for coated spheres~\cite{Mie1908,BohrHuff1983,Li1994}, with the dyadic Green's tensor built from the exact scattering amplitudes as detailed in Appendix~\ref{app:mie}. The multipole expansion is converged at $n_\mathrm{max} = 60$ for the fixed center-to-dipole distance $R = 23$~nm.

\section{\label{sec:results}Results}

\subsection{\label{sec:kernel comparison}Coupling kernel spectrum}

\begin{figure}[t]
 \includegraphics[width=\columnwidth]{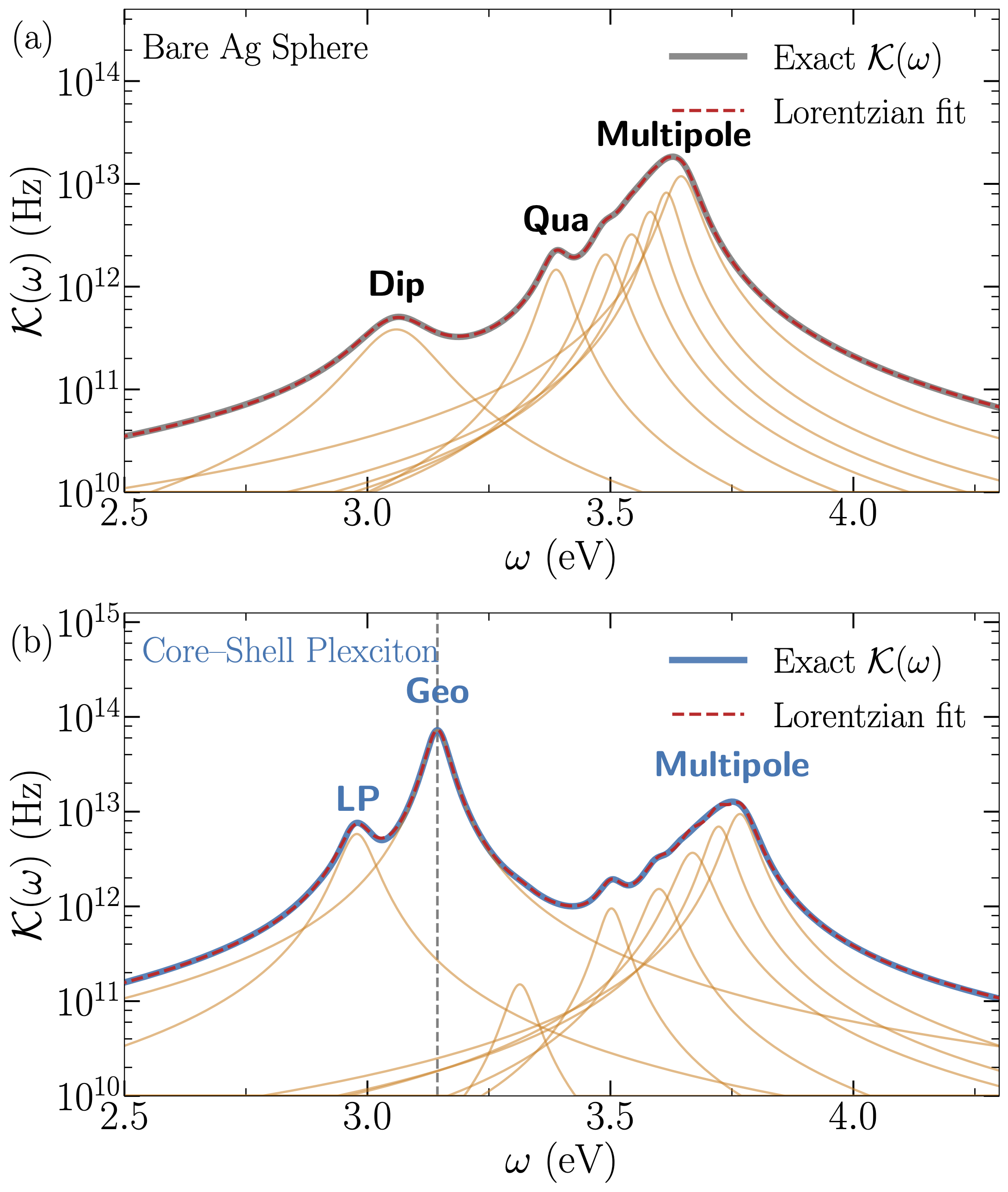}
\caption{\label{fig:kernel}Kernel spectrum $\mathcal{K}(\omega)$ on a logarithmic scale for a $z$-oriented quantum dot ($d_{eg}=24$~D) placed 
at a fixed metal-to-emitter gap of $d=3$~nm. (a) Spectrum for the bare Ag nanosphere of radius $a=20$~nm. (b) Spectrum for the Ag/J-aggregate core--shell plexciton with the same core radius and a shell thickness of $h=2$~nm. Thick solid curves (gray for the bare sphere, blue for the plexciton) represent the Mie theory calculation. Dashed red lines denote the total multi-Lorentzian fit defined in Eq.~(\ref{eq:kernel lorentzian}). Solid orange lines show the individual Lorentzian components comprising the pseudo-mode expansion. A vertical dashed line marks the frequency of the geometric mode (Geo) at 3.14~eV.}
\end{figure}

Figure~\ref{fig:kernel} compares $\mathcal{K}(\omega)$ for the bare and coated nanospheres. The kernel for the bare sphere exhibits a hierarchy of multipole contributions that merge into a broad spectral envelope, as shown in Fig.~\ref{fig:kernel}(a). For a 3~nm gap, higher-order multipoles are enhanced by the near-field weighting. The dipolar localized surface plasmon resonance (3.06 eV) carries modest spectral weight and appears as a lower-energy shoulder. Resolved quadrupolar and octupolar contributions emerge \notablue{around 3.45 eV}, while a dominant multipole complex centered near 3.65~eV carries most of the total spectral weight. These features are consistent with previous results in the literature~\cite{Vanvlack2012}. Seven Lorentzians are used to \notablue{accurately fit the} kernel spectrum, all with positive amplitudes. This positivity is consistent with the fact that each multipole order $\ell$ contributes independently to the imaginary part of the Green's tensor without cross-channel interference \cite{CuarteroGonzalez2020}.

Fig.~\ref{fig:kernel}(b) shows the \notablue{kernel} spectrum for the coated configuration. Eight Lorentzians with positive amplitudes are required to reproduce the fully converged spectrum. Comparing the two decompositions peak by peak reveals how the molecular shell transforms the spectral landscape. Lorentzian fit parameters are given in Appendix \ref{app:fit}.

The dipolar contribution \notablue{in Fig. \ref{fig:kernel}(b) redshifts relative to the bare sphere [Fig. \ref{fig:kernel}(a)], forming a lower polariton peak} near 2.98~eV. According to the plasmon hybridization picture~\cite{Antosiewicz2014}, this mode corresponds to the bonding combination of the plasmon and exciton, with strong surface charge density at both the inner and outer shell interfaces. This hybridization-enhanced charge redistribution accounts for the nearly sevenfold increase in near-field spectral weight relative to the bare sphere dipolar mode. \notablue{The single-Lorentzian coupling criterion using the $j=1$ fitting parameters gives $2A_1/B_1^2 \approx 0.66$ , for $A_1=440$ meV$^2$ and $B_1= 37$ meV (see Appendix \ref{app:fit}). Single-mode weak coupling with a dipole emitter is therefore expected at this frequency. However, it is shown below that strong coupling is possible at the dipole-to-sphere distance considered, as evidenced by the emergence of a Rabi splitting in the near-field spectrum. This apparent contradiction highlights the limitations of single-mode concepts to describe broadband optical nanocavities. }

\begin{figure*}[t]
 \includegraphics[width=0.8\textwidth]{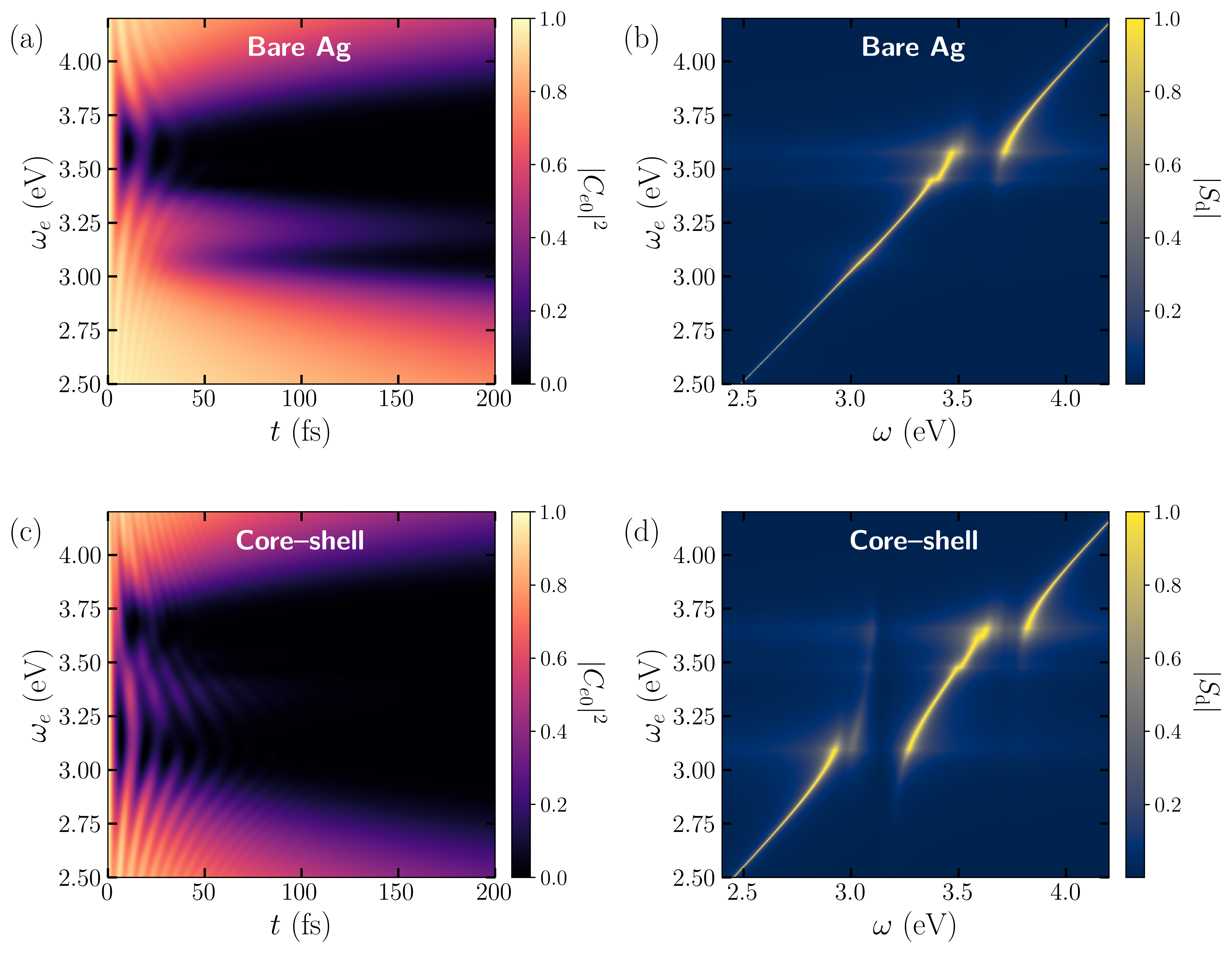}
\caption{\label{fig:spectral_maps}Broadband emitter dynamics and coherence spectra as a function of transition frequency $\omega_e$ at a fixed gap of 3~nm. (a) Excited-state population dynamics $|C_{e0}(t)|^2$ for the bare silver nanosphere at different dipole frequencies $\omega_e$. (b) Normalized excited-state coherence spectra for the bare sphere, exhibiting a single avoided crossing near the Near-UV multipole at 3.65~eV. (c) Population dynamics $|C_{e0}(t)|^2$ for the coated core--shell system at different dipole frequencies $\omega_e$. (d) Normalized coherence spectra for the core--shell system. The molecular coating induces two distinct regions of coherent exchange: a blue-shifted multipole near 3.75~eV, and a merged multi-branch avoided crossing in the visible spectrum.}
\end{figure*}

The geometric mode emerges as the dominant peak in the spectrum at 3.14~eV, in between the polariton branches. The exciton linewidth sets the FWHM to $\sim50$~meV rather than the broad metallic damping. Because the emitter is placed only 1~nm from the outer shell surface, it couples intensely to a manifold of high-order multipole fields that constructively sum at this frequency. \notablue{From a single-Lorentzian perspective, taking the dominant spectral weight at that frequency ($j=2$), the coupling criterion gives $2A_2/B_2^2= 12.1$ ($A_2 = 3717~\text{meV}^2$ and $B_2 = 25$ meV) for the geometric mode, thus strong coupling is expected for a quantum dot dipole at this frequency. This expectation is confirmed and refined later from a multi-mode analysis of the kernel poles.} 

The intermediate multipole region between 3.3 and 3.6~eV is strongly suppressed in the coated system. These modes are detuned from the exciton and do not hybridize with it directly. Instead, the negative permittivity window of the shell repels them, acting as a spectral barrier that extends from just above the exciton frequency to roughly 3.4~eV.

The dominant multipole complex centered near 3.65~eV in the bare sphere shifts upward to the 3.72--3.77~eV range in the coated system. This displacement cannot be explained by a simple two-oscillator polariton model because these high-order modes are far detuned from the exciton frequency. The region instead merges contributions from the upper polariton with the modified multipole tail pushed upward by the negative permittivity window. \notablue{Despite being  composed of multiple Lorentzians, a single-Lorentzian analysis using the largest fit contribution ($j=7$) gives $2A_7/B_7^2=1.41$ at 3.64 eV for a bare silver sphere ($A_7 = 854~\text{meV}^2$ and $B_7 = 35$ meV), which is consistent with strong coupling at these dipole-to-sphere distances (first studied in Ref. \cite{Vanvlack2012}). It is shown below that the core-shell structure preserves strong coupling conditions.} 

\begin{figure*}[t]
  \includegraphics[width=\textwidth]{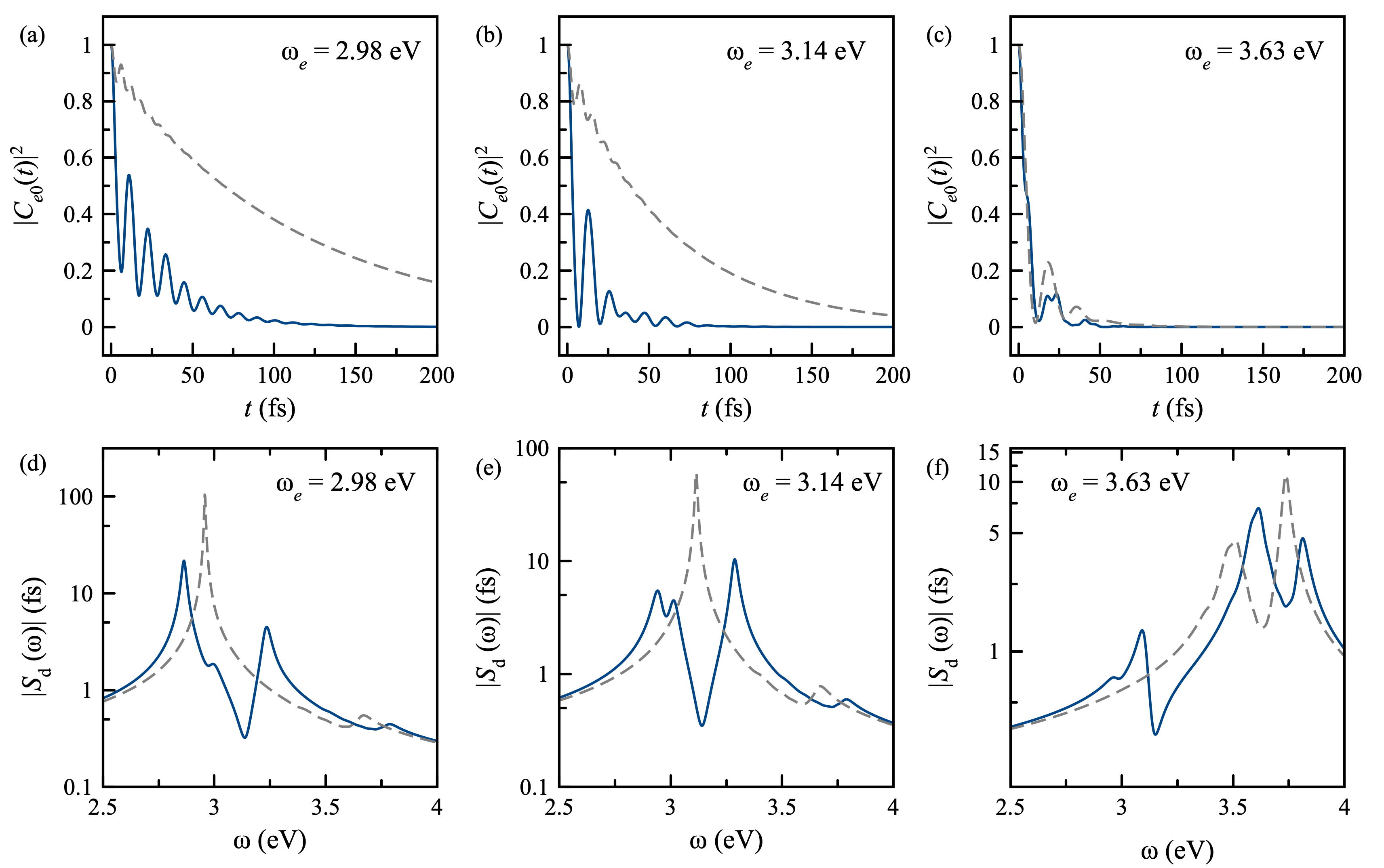}
\caption{\label{fig:new_channels}\notablue{(a) Excited state population dynamics for a quantum emitter with frequency $\omega_e=2.98$ eV, for a bare silver sphere (dashed line) and the core--shell structure (solid line); (b) The same as panel (a) for $\omega_e= 3.14$ eV; (c) The same as panel (a) for $\omega_e=3.63$ eV; (d) Coherence spectrum $|S_{\rm d}(\omega)|$ corresponding to the qubit dipole evolution shown in panel (a); (e) Coherence spectrum corresponding to panel (b); (f) Coherence spectrum corresponding to panel (c). In all panels the metal-to-emitter distance is $d=3$ nm, the sphere radius $a= 20$ nm, and the emitter dipole $d_{eg}=24$ D.} }
\end{figure*}

\subsection{\label{sec:dynamics}Quantum Emitter dynamics}

\notablue{Figure~\ref{fig:spectral_maps} compares the excited dipole population dynamics $|C_{e0}(t)|^2$ at different qubit dipole frequencies $\omega_e$ and the corresponding coherence spectra $\mathrm{Im}\,C_{e0}(\omega)$, for a quantum dot dipole interacting with a bare silver sphere [Figs.~\ref{fig:spectral_maps}(a) and \ref{fig:spectral_maps}(b)] and core-shell structure [Figs.~\ref{fig:spectral_maps}(c) and \ref{fig:spectral_maps}(d)]. The molecular coating is shown to reshape the electromagnetic vacuum across the entire spectral range}. 

For the bare sphere, Fig.~\ref{fig:spectral_maps}(a) shows two distinct dynamical regions. Above approximately 3.4~eV, clear oscillatory fringes indicate coherent population exchange with the near-UV multipole modes, with the tightest fringe spacing near 3.6~eV where the spectral weight is largest. Below 3.4~eV, the population decays monotonically without revivals, transitioning from moderate Purcell-enhanced decay near the dipolar LSPR at 3.06~eV to negligible coupling at lower frequencies. Figure~\ref{fig:spectral_maps}(b) shows a single avoided crossing near 3.6~eV, while below 3.4~eV the spectrum follows the emitter frequency, consistent with previous studies of emitters near bare metallic surfaces~\cite{Delga2014}.

The coated system presents a qualitatively different picture. The population map [Fig.~\ref{fig:spectral_maps}(c)] shows coherent 
oscillations across a much broader range of frequencies. From 2.8 to 3.3~eV, pronounced oscillatory fringes appear where the 
bare sphere showed only monotonic decay. The tightest fringe spacing occurs near the geometric mode at 3.14~eV, where the population collapses within approximately 10~fs. A second region of oscillations persists near 3.7--3.8~eV. Between these two regions, from roughly 3.3 to 3.6~eV, the population decays without revivals, reflecting the suppressed spectral density in the $\mathrm{Re}[\epsilon_\mathrm{sh}]<0$ spectral gap. The coherence map [Fig.~\ref{fig:spectral_maps}(d)] reveals the spectral origin of this behavior. Two distinct regions of coherent exchange emerge. Near 3.75~eV, an isolated anti-crossing appears with well-separated upper and lower branches. The visible range from roughly 2.8 to 3.3~eV is qualitatively different. When an emitter sweeps through a single well-isolated near-field resonance, the coherence spectrum shows an avoided crossing~\cite{Tan2021}. If the geometric mode and the lower polariton acted as independent near-field channels, each would produce its own avoided crossing centered at 2.98~eV and 3.14~eV respectively. Instead, the spectral branches merge into a continuous multi-branch structure, indicating that these two plexcitonic features overlap in the near field to form a coupled multi-mode continuum as experienced by the emitter.

\begin{table*}[t]
\caption{\notablue{Frequency, decay, and amplitude parameters $(\omega_j,\gamma_j, |R_j|)$ of the dominant pole contributions to the dipole coherence evolution $C_{e0}(t)=\sum_jR_j{\rm exp}[(i\omega_j-\gamma_j)t]$, for a dipole interacting with the bare silver sphere and the  molecular aggregate core-shell structure. Different qubit frequencies $\omega_e$ are shown. The estimated Rabi splitting from the pole frequencies $\Omega_R^{\rm SM}$ is compared with the splitting in the coherence spectrum $\Omega_R$. SM-SC: single-mode strong coupling; WC: weak coupling; MM-SC: multi-mode strong coupling. $d=3$ nm, $a= 20$ nm, $d_{eg}=24$ D.}}
\label{tab:pole_analysis}
\begin{ruledtabular}
\begin{tabular}{llcccccc}
$\omega_e$ (eV) & System & $\omega_j-\omega_e$ (meV) & $|R_1|$, $|R_2|$, $|R_3|$ & $\gamma_1, \gamma_2, \gamma_3$ (meV) & $\Omega_R^{\rm poles}$ (meV) & $\Omega_R$ (meV) & Regime \\ 
\colrule
2.98 & Bare Ag     & -23, 82, 675  & 0.96, 0.02, 0.02 & 3, 81, 34 & 0 & 0 & WC \\
      & Core-shell & -117,  13, 253  & 0.67, 0.07, 0.24  &  10, 33, 19 & 370 & 372 & SM-SC \\ 
\colrule
3.14 & Bare Ag    &   -83, -25, 519 &0.04, 0.96, 0.03  & 79, 5, 33 & 0 & 0 & WC \\
      & Core-shell  & -197, -126, 147 & 0.29, 0.25, 0.45 & 24, 24, 14 & 344  & 348 & MM-SC \\       
\colrule
      3.63 & Bare Ag     & -151, -113, 106 & 0.14, 0.21, 0.52 & 29, 26, 16 & 229 & 227 & MM-SC \\
              & Core-shell  & -43, - 8, 178 & 0.22, 0.33, 0.27 & 28, 25, 23 & 221  & 201 &  MM-SC  
\end{tabular}
\end{ruledtabular}
\end{table*}

\subsubsection{\notablue{Kernel pole analysis: multi-mode vs single-mode coupling}}

\notablue{Figure~(\ref{fig:new_channels}) shows cuts of the population and spectral density maps in Fig.~\ref{fig:spectral_maps} at selected qubit frequencies: $\omega_e = 2.98$ eV (lower polariton),  $3.14$ eV (geometric mode) and $3.63$ eV (multipole). Table \ref{tab:pole_analysis} shows the kernel pole parameters $s_m$ that contribute with the largest amplitudes $R_m$ to the excited state amplitude in Eq. (\ref{eq:residues}). }

At $\omega_e = 2.98$~eV, the bare sphere interacts only with the weak tail of the dipolar LSPR, with a Purcell factor $F_P \approx 551$. \notablue{Table \ref{tab:pole_analysis} shows that a single pole carries 96\% of the residue weight with oscillation frequency $\omega_1-\omega_e=-23 $meV and decay constant $\gamma_1= 3$ meV.  Other contributions have much smaller amplitudes and decay much faster. This leads to a single-mode low-$Q$ coherence oscillation, where $Q\equiv (\omega_1-\omega_e)/\Gamma= 23 {\rm meV}/83{\rm meV}\approx 0.3$ ($\Gamma=B_1$ from Table \ref{tab:fit bare}). This is consistent with the absence of Rabi splitting in the spectrum (weak coupling). The coated system drives the emitter into the strong coupling regime,  with the dominant contribution of two poles ($|R_1|=0.67$, $|R_3|=0.24$), with a third pole contributing only 7\%. From the relative frequency between the two dominant poles, we can estimate a Rabi splitting $\Omega_R^{\rm poles}=370$ meV, which closely agrees with the Rabi splitting extracted from  the coherence spectrum in Fig. \ref{fig:new_channels}(d). The contribution of the two poles gives damped coherence oscillation with $Q= (\omega_3-\omega_1)/\Gamma \approx 4.5$, which can be considered as single-mode strong coupling.}

\notablue{At $\omega_e = 3.14$~eV the emitter is tuned to the geometric mode of the coated particle, while for the bare sphere this frequency lies on the blue tail of the dipolar LSPR ($F_P \approx 578$). Table~\ref{tab:pole_analysis}
shows that the bare sphere again responds through a single pole carrying 96\% of the residue weight, at $\omega_2-\omega_e=-25$~meV with $\gamma_2=5$~meV, the remaining 7\% being shared by the broad LSPR pole ($-83$~meV, $\gamma=79$~meV) and the far-detuned multipole manifold ($+519$~meV). The coherence oscillation
is overdamped, $Q=25~{\rm meV}/83~{\rm meV}\approx0.3$, and the population decays monotonically [Fig.~\ref{fig:new_channels}(b)], leaving a single narrow peak in the coherence spectrum with no Rabi splitting (weak coupling). The core-shell system responds instead through three poles of comparable weight $|R_j|$, resulting in a multi-frequency coherence oscillation pattern with a triplet feature in the coherence spectrum. From the outermost pole frequencies we estimate $\Omega_R^{\rm poles}=(\omega_3-\omega_1)\approx 344$~meV, which closely agrees with the spectral splitting $\Omega_R=348$~meV  seen in the spectrum Fig.~\ref{fig:new_channels}(e). This multi-pole oscillatory regime can be classified as multi-mode strong coupling.}

\notablue{At $\omega_e = 3.63$~eV the emitter is tuned to the multipole resonance of the bare silver sphere, the only frequency region in which the bare sphere can give strong coupling at dipole-to-sphere distances considered. For the bare sphere, the blue-detuned pole at $\omega_3=\omega_e+106\,{\rm meV}$ is more dominant ($52\%$), but there is no clear separation of magnitudes from other pole contributions. From the two most dominant poles we have  $\Omega_R^{\rm poles}=(\omega_3-\omega_2)= 229$~meV, which is close to the splitting shown in the spectrum of Fig. \ref{fig:new_channels}(f). For the core-shell system at this frequency, the pole contributions are more balanced, and the estimated Rabi splitting is $\Omega_R^{\rm poles}=(\omega_3-\omega_1)= 221$~meV, which overestimates the splitting $\Omega_R= 201$ meV in Fig.~\ref{fig:new_channels}(f). The geometric mode resonance feature is also present in the spectrum, despite the large detuning with the qubit dipole ($\sim$490~meV).}

\subsubsection{\notablue{Distance dependence of geometric-mode strong coupling}}
\begin{figure}
\includegraphics[width=0.5\textwidth]{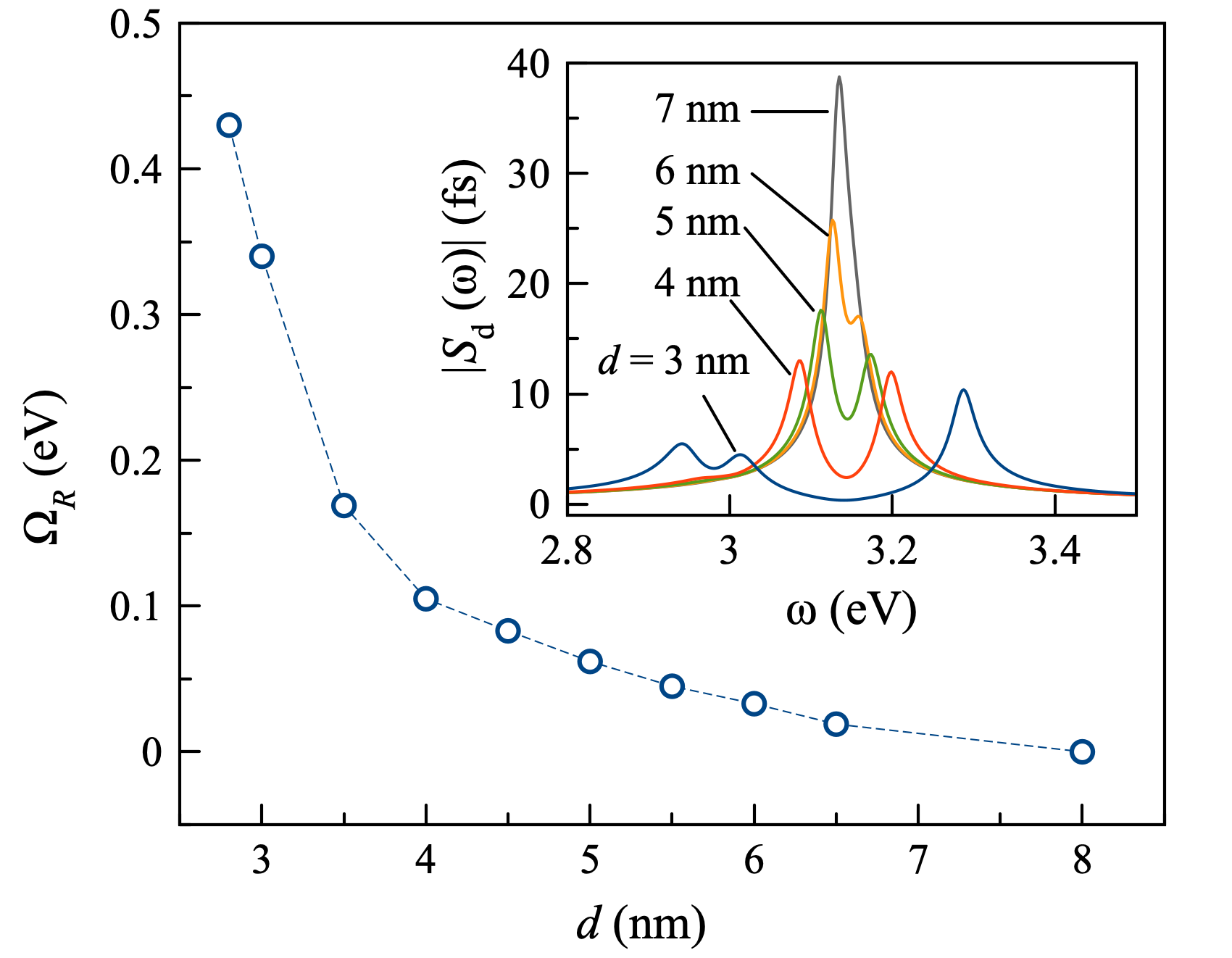}
\caption{\notablue{Rabi splitting $\Omega_R$ as a function of dipole-to-silver distance $d$, extracted from the dipole coherence spectrum for $\omega_e=3.14$ eV. {\it Inset}: coherence spectrum for selected separation distances.}}
\label{fig:distance dependence}
\end{figure}

\notablue{The coupling strength to the geometric mode depends on the distance from the emitter to the molecular aggregate surface. Figure~\ref{fig:distance  dependence} shows the coherence spectrum at $\omega_e = 3.14$~eV for different metal-to-emitter distances. As discussed above, for $d=3$ nm the coupled spectrum exhibits a triplet with the outer branches split by 348~meV. As the emitter is moved away from the molecular aggregate shell surface, the spectrum turns into a doublet that eventually merges into a single resonance at the qubit frequency.  The splitting decays rapidly and can be fitted to an exponential function $\Omega_R(z)\sim {\rm exp}[-z/b]$, with $1/e$ decay length $b\approx1$ nm. Weak coupling at the geometric mode is expected for dipoles separated from the nearest aggregate shell surface by about $\sim 3$ nm or more. This strong distance dependence is consistent with the behavior of localized plasmon resonances in other nanocavity systems, which require small dipole-to-sphere separations \cite{Vanvlack2012,Delga2014} or nanocavity gap sizes \cite{Chikkaraddy2016,SaezBlazquez2022} for strong coupling to occur.}

\section{Conclusions and Outlook}
\label{sec:conclusions}

We adopted a recently developed Lorentzian approximation method for mapping the non-Markovian light-matter interaction kernel that describes the electronic dynamics of  a quantum emitter interacting with an optical nanocavity field. The method decomposes the coupling kernel into a set of Lorentzian pseudo-modes, from which closed-form analytical expressions for the excited-state population and the emitted photon coherence can be derived \cite{Rema2025}. The method is particularly useful for modeling complex nanophotonic environments where the local density of optical states exhibits sharp, highly structured spectral features, including mode-coupled phenomena such as Fano resonances. 

It is shown that by coating a silver metal sphere with a thin molecular J-aggregate layer (core-shell structure), the  near-field electromagnetic vacuum of the nanoparticle changes significantly such that a quantum emitter that would only undergo Purcell-enhanced spontaneous emission without coating, becomes strongly coupled and experiences Rabi oscillations with a multi-mode splitting spectrum. The nanoparticle-emitter distance remains unchanged. This weak-coupling to strong-coupling crossover occurs due to plasmon-exciton hybridization within the core-shell structure \cite{Antosiewicz2014},  which spectrally modifies the local density of optical states. The negative real part of the J-aggregate permittivity gives metallic character to the molecular shell enabling the formation of geometric resonances to which quantum emitters can strongly couple, over a frequency region where the bare metal nanoparticle does not support strong coupling.  

The system studied here is within reach of current technology. Conformal J-aggregate shells with the required oscillator strengths are routinely deposited around colloidal nanoparticles at room temperature~\cite{Fofang2008,Fofang2011,Schlather2013,Zengin2015}. DNA origami scaffolds~\cite{Kuzyk2012} and cucurbituril molecular spacers~\cite{Chikkaraddy2016} provide the sub-nanometer positional control needed to place individual solid-state emitters at prescribed distances. The predicted population beatings occur on timescales of $\sim 10$~fs, accessible to ultrafast pump-probe and time-resolved measurements~\cite{Wersall2017,Park2019,Yang2026}. The observation of the predicted triplet spectral feature spanning $\sim 350$~meV due to strong coupling with plasmon-exciton modes at visible frequencies should be straightforward. 

Molecular aggregate coating layers can be used to modify the electromagnetic vacuum for other nanoparticle geometries, such as nanoparticle-on-mirror configurations, plasmonic dimers, and tip-enhanced geometries. The theoretical framework we used is completely general and can be applied for studying arbitrary nanocavity configurations provided that the Green's dyadic tensor is known. Molecular coatings can therefore become valuable design tools for engineering photonic reservoirs of optical nanocavities to enable new types of chemical reactions \cite{Herrera2016,Wellnitz2021,Ying2026}, nanoscale energy transfer platforms \cite{Buendia2024} and nanoscale quantum control schemes \cite{Bedingfield2025}.

\begin{acknowledgments}
We thank Tomasz Antosiewicz for sharing data from Ref.~\cite{Antosiewicz2014} for validation. A.S.R. received support from USACH Excellence Graduate Scholarship. A.E.R.L. is supported by ANID-Fondecyt Iniciaci\'on No. 11250638. F.H. is supported by ANID-Fondecyt Regular Grant No. 1262603. This work was also supported by the Air Force Office of Scientific Research under Award No. FA9550-22-1-0245 and ANID-Millennium Science Initiative Program No. ICN17\_012.
\end{acknowledgments}

\section*{Disclosure statement}
No potential competing interest was reported by the authors.

\onecolumngrid


\appendix

\section{\label{app:mie}Lorenz--Mie theory and the 
Green's tensor}

This appendix outlines the derivation of the kernel spectrum $\mathcal{K}(\omega)$ from classical Mie scattering theory. We construct the dyadic Green's tensor at the emitter position directly from the exact Mie coefficients of the coated sphere, which yields the coupling kernel via Eq.~(\ref{eq:g coupling}).

\subsection{\label{app:mie coated}Mie coefficients for the coated sphere}
To avoid notational conflict with the standard transverse magnetic Mie coefficients $a_n$, in this appendix we denote the radius of the metallic core as $R_1$ (corresponding to $a$ in the main text). The core is coated with a concentric shell of thickness $h$, yielding an outer radius of $R_2 = R_1 + h$. The entire structure is embedded in an unbounded medium. The three regions have permittivities $\epsilon_j$ and wavenumbers $k_j = \omega\sqrt{\epsilon_j}/c$ ($j = 1,2,3$ for core, shell, exterior); all media are non-magnetic. The Lorenz--Mie scattering coefficients for this geometry are~\cite{BohrHuff1983}
\begin{equation}\label{eq:an coated}
  a_n = \frac{
    m_2\,\psi_n'(k_3 R_2)
    - \psi_n(k_3 R_2)\,\mathcal{A}_n
  }{
    m_2\,\xi_n'(k_3 R_2)
    - \xi_n(k_3 R_2)\,\mathcal{A}_n
  }\,,
\end{equation}
where $m_j = k_j/k_3$ is the relative refractive index of 
region $j$, and $\psi_n(x) = x\,j_n(x)$, 
$\xi_n(x) = x\,h_n^{(1)}(x)$ are Riccati--Bessel 
functions. The effective logarithmic derivative 
$\mathcal{A}_n \equiv \mathcal{A}_n(m_2 k_3 R_2)$ 
incorporates the effect of the core:
\begin{equation}\label{eq:an eff}
  \mathcal{A}_n
  = \frac{
    \psi_n'(m_2 k_3 R_2)
    - A_n^c\,\chi_n'(m_2 k_3 R_2)
  }{
    \psi_n(m_2 k_3 R_2)
    - A_n^c\,\chi_n(m_2 k_3 R_2)
  }\,,
\end{equation}
with $\chi_n(x) = -x\,y_n(x)$ and the core ratio
\begin{equation}\label{eq:an core}
  A_n^c = \frac{
    m_2\,\psi_n(m_2 k_3 R_1)\,\psi_n'(m_1 k_3 R_1)
    - m_1\,\psi_n'(m_2 k_3 R_1)\,\psi_n(m_1 k_3 R_1)
  }{
    m_2\,\chi_n(m_2 k_3 R_1)\,\psi_n'(m_1 k_3 R_1)
    - m_1\,\chi_n'(m_2 k_3 R_1)\,\psi_n(m_1 k_3 R_1)
  }\,.
\end{equation}
The TE (magnetic) Mie coefficient $b_n$ follows from the 
same expressions with the replacement 
$m_j \to 1/m_j$~\cite{BohrHuff1983}. Far-field cross 
sections are
\begin{align}
  \sigma_\mathrm{ext} &= \frac{2\pi}{k_3^2}
  \sum_{n=1}^\infty(2n+1)\,\mathrm{Re}[a_n+b_n]\,,
  \label{eq:sigma ext}\\
  \sigma_\mathrm{sca} &= \frac{2\pi}{k_3^2}
  \sum_{n=1}^\infty(2n+1)\,
  (|a_n|^2+|b_n|^2)\,.
  \label{eq:sigma sca}
\end{align}

\subsection{\label{app:gzz}Green's tensor and the 
kernel spectrum}

The electromagnetic dyadic Green's tensor for the multilayered sphere can be constructed using the scattering superposition method~\cite{Li1994,Tai1971}. The total tensor is separated into a free-space component and a scattering correction. The scattering correction is expanded in vector spherical harmonics, with coefficients determined by boundary matching at each interface. When both the source and the 
field are located in the exterior region, the transverse magnetic (TM) and transverse electric (TE) scattering amplitudes of the Green's tensor 
reduce directly to the negative of the standard Mie coefficients, $-a_n(\omega)$ and $-b_n(\omega)$, respectively~\cite{Arruda2017}. 
This exact equivalence forms the central link between macroscopic scattering theory and the local electromagnetic environment experienced by the quantum emitter.

For a $z$-oriented (radial) dipole located on the symmetry axis at a distance $R = R_1 + d$ from the sphere center, only the $m = 0$, TM terms of the multipole expansion survive~\cite{Li1994,Arruda2017}. The scattering part of $G_{zz}$ at the coincident point $\mathbf{r} = \mathbf{r}' = R\,\hat{z}$ simplifies to:
\begin{equation}\label{eq:gzz mie}
  G_{s,zz}(R,R)\big|_{\theta=0}
  = -\frac{ik_3}{4\pi}\sum_{n=1}^\infty
  (2n+1)\frac{n(n+1)}{(k_3 R)^2}\,
  a_n\bigl[h_n^{(1)}(k_3 R)\bigr]^2.
\end{equation}
The total Purcell factor for a radially oriented dipole follows from the imaginary part of the full (free-space plus scattering) Green's tensor~\cite{Arruda2017}:
\begin{equation}\label{eq:purcell perp}
  F_P=\frac{\Gamma_\perp}{\Gamma_0} = 1
  - \frac{3}{2}\sum_{n=1}^\infty n(n+1)(2n+1)\,
  \mathrm{Re}\!\left[
  a_n\!\left(\frac{h_n^{(1)}(k_3 R)}{k_3 R}\right)^{\!2}
  \right].
\end{equation}
This classical Purcell enhancement is proportional to the projected local density of optical states (LDOS), providing the direct physical link between the macroscopic antenna properties and the quantum memory kernel. The kernel spectrum entering the IDE~(\ref{eq:Ce0}) is therefore defined directly from this exact Green's tensor as:
\begin{equation}\label{eq:kernel from gzz}
  \mathcal{K}(\omega) = \frac{d_{eg}^2\omega^2}
  {\pi\hbar\epsilon_0 c^2}\,
  \Im\!\left[G_{s,zz}(R,R;\omega)\right],
\end{equation}
which is evaluated numerically by summing Eq.~(\ref{eq:gzz mie}) over multipole orders $n = 1,\ldots,n_\mathrm{max}$. We use $n_\mathrm{max} = 60$, which provides convergence at the extreme near-field fixed center-to-center distance $R = a+d = 23$~nm.

\section{\label{app:fit}Lorentzian fit parameters}

Tables~\ref{tab:fit bare} and~\ref{tab:fit coat} list the parameters of the multi-Lorentzian expansion [Eq.~(\ref{eq:kernel lorentzian})] used to fit the kernel spectrum $\mathcal{K}(\omega)$ for each configuration. 

For the bare Ag sphere, seven Lorentzians are needed, all with positive amplitudes. The lowest-frequency term ($j=1$, $\Omega_1\approx 3.06$~eV) corresponds to the dipolar LSPR; terms $j=2$--$5$ capture the resolved quadrupolar, octupolar, and higher multipole contributions 
in the 3.39--3.55~eV range; and the dominant multipole complex near 3.65~eV is described by terms $j=6$--$7$. 
All amplitudes are positive, consistent with the absence of cross-channel interference: each multipole order $\ell$ contributes independently to 
$\mathrm{Im}[G_{zz}]$.

For the core--shell system ($f=0.3$), eight Lorentzians are needed to capture the fully converged spectrum. The lower polariton is captured primarily by term $j=1$ near 2.98~eV. The geometric mode at $\omega_\mathrm{geo}\approx 3.14$~eV appears as the single dominant term ($j=2$), carrying the largest spectral weight in the entire fit. An isolated intermediate multipole ($j=3$) and the 
upper-polariton/multipole complex ($j=4$--$8$) complete the decomposition, all with positive amplitudes.

\begin{table}[h]
\caption{\label{tab:fit bare}Lorentzian fit parameters for the bare Ag sphere ($a = 20$~nm, $d = 3$~nm). $n = 7$ terms, all with positive amplitudes.}
\begin{ruledtabular}
\begin{tabular}{cccc}
$j$ & $\Omega_j$ (eV) & $B_j$ (eV)    & $A_j$ ($\text{meV}^2$) \\
\hline
1 & 3.0596 & 0.0828 &      65.6 \\
2 & 3.3883 & 0.0321 &      97.2 \\
3 & 3.4903 & 0.0336 &     142.7 \\
4 & 3.5435 & 0.0319 &     212.7 \\
5 & 3.5820 & 0.0282 &     311.9 \\
6 & 3.6151 & 0.0273 &     463.9 \\
7 & 3.6455 & 0.0348 &     854.1 \\
\end{tabular}
\end{ruledtabular}
\end{table}

\begin{table}[h]
\caption{\label{tab:fit coat}Lorentzian fit parameters for the Ag/J-aggregate core--shell ($a = 20$~nm, $h = 2$~nm, $f = 0.3$, $d = 3$~nm). The fit utilizes $n = 8$ positive-amplitude terms. The geometric mode ($j=2$) dominates the visible spectrum.}
\begin{ruledtabular}
\begin{tabular}{cccc}
$j$ & $\Omega_j$ (eV) & $B_j$ (eV)    & $A_j$ ($\text{meV}^2$) \\
\hline
 1 & 2.9782 & 0.0366 &     439.5 \\
 2 & 3.1441 & 0.0248 &    3717.2 \\
 3 & 3.3136 & 0.0305 &       9.5 \\
 4 & 3.5022 & 0.0275 &      54.2 \\
 5 & 3.5999 & 0.0353 &     111.6 \\
 6 & 3.6690 & 0.0371 &     282.3 \\
 7 & 3.7228 & 0.0298 &     428.8 \\
 8 & 3.7665 & 0.0320 &     625.4 \\
\end{tabular}
\end{ruledtabular}
\end{table}

\section{Integro-differential equation and \notablue{near-field} spectrum}
\label{app:ide}

This appendix outlines the derivation of the integro-differential equation for the excited-state amplitude [Eq.~(\ref{eq:Ce0})], \notablue{the formal definition of the near-field dipole coherence spectrum,} and the stationary single-photon spectrum used to define the strong coupling criterion $2A_j/B_j^2 > 1$~\cite{Rema2025}.

Inserting the Wigner--Weisskopf ansatz [Eq.~(\ref{eq:wf ansatz})] into the time-dependent Schr\"odinger equation $i\hbar\partial_t\lvert\psi\rangle = \hat{\mathcal{H}}\lvert\psi\rangle$ and projecting onto the basis states yields a system of coupled differential equations for the amplitudes. Formally integrating the single-photon amplitude $C_{g1}(\omega,t)$ with the initial condition $C_{g1}(\omega,0)=0$ yields
\begin{equation}\label{eq:app cg1 formal}
  C_{g1}(\omega,t) = -i\int_0^t dt'\,
  \sqrt{\mathcal{K}(\omega)}\,C_{e0}(t')\,
  \mathrm{e}^{i(\omega-\omega_e)t'}\,,
\end{equation}
where $\sqrt{\mathcal{K}(\omega)} = d_{eg}\,|g(\omega)|/\hbar$. Substituting this formal solution back into the equation of motion for $C_{e0}(t)$ directly yields the integro-differential equation
\begin{equation}\label{eq:app ide final}
  \dot{C}_{e0}(t) = -\int_0^t dt'\,
  \mathcal{K}(t-t')\,C_{e0}(t')\,,
\end{equation}
with the memory kernel $\mathcal{K}(t-t') = \int_0^\infty d\omega\,\mathcal{K}(\omega)\,\mathrm{e}^{-i(\omega-\omega_e)(t-t')}$. This establishes Eqs.~(\ref{eq:Ce0}) and~(\ref{eq:delay kernel}) of the main text.

\notablue{To evaluate the spectral features of the near-field emission, we separate the real and imaginary parts of the fast-oscillating dipole amplitude $\tilde{C}_{e0}(t) \equiv C_{e0}(t)\mathrm{e}^{-i\omega_e t}= \tilde{C}_{e0}'(t) + i\tilde{C}_{e0}''(t)$.  The dipole coherence fully captures the phase response about the coupled dipole system. We define the near-field dipole coherence spectrum $S_{\rm d}(\omega)$ in terms of the imaginary part as }
\begin{equation}\label{eq:app coherence}
  S_{\rm d}(\omega) = \int_{-\infty}^{\infty} dt\, \tilde{C}_{e0}''(t)\, \mathrm{e}^{-i\omega t}.
\end{equation}

\begin{figure}[t]
  \centering
  \includegraphics[width=0.7\textwidth]{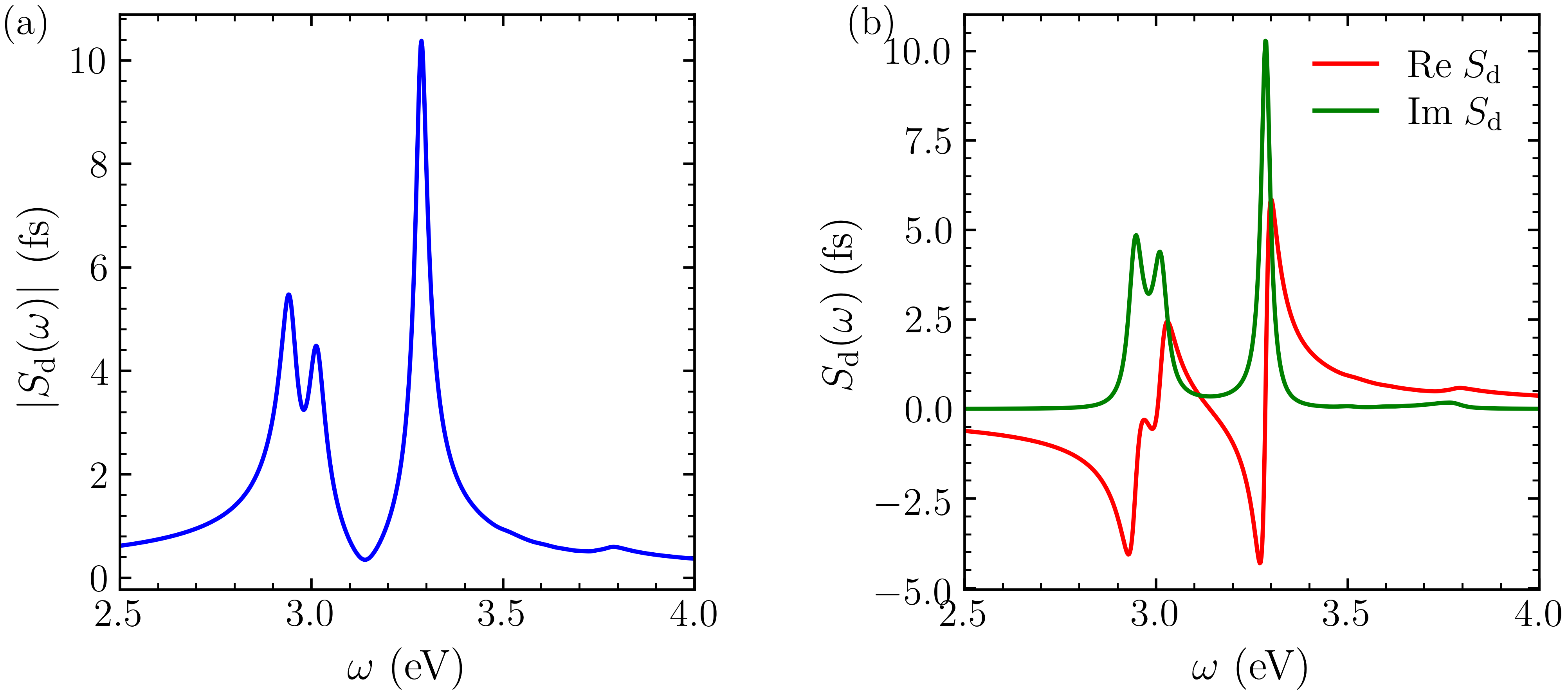}
  \caption{\label{fig:app_spectrum} Near-field dipole coherence spectrum $S_d(\omega)$ for a quantum emitter at a distance $d=3$~nm tuned to the geometric mode ($\omega_e = 3.14$~eV), corresponding to the spectrum shown in Fig.~\ref{fig:new_channels}(e) of the main text. (a) Absolute magnitude $|S_d(\omega)|$, showing the splitting feature of strong coupling; (b) The corresponding real and imaginary parts of the spectrum, which preserve the phase information of the near-field dipole dynamics.}
\end{figure}

\notablue{Figure~\ref{fig:app_spectrum}(a) shows the magnitude of the coherence spectrum $|S_{\rm d}(\omega)|$, which is the quantity reported in the main text, for a quantum emitter at a distance $d=3$~nm from the sphere surface, tuned to the geometric mode ($\omega_e = 3.14$~eV). This is the spectrum shown in Fig.~\ref{fig:new_channels}(e) (core-shell). Figure ~\ref{fig:app_spectrum}(b) shows that the real and imaginary parts of $S_{\rm d}(\omega)$ contain the same resonance peak positions as the absolute magnitude, but retaining the phase information of the near-field emission. Phase information can in principle be measured using heterodyne near-field microscopy techniques~\cite{Raschke2023,Esses2026}.}

The calculation of far-field emission patterns is beyond the scope of this work, but the stationary single-photon spectrum $|C_{g1}^\infty(\omega)|^2$ would in principle determine the far-field spectral features. Taking the limit $t\to\infty$ in Eq.~(\ref{eq:app cg1 formal}) produces an integral that is exactly the Laplace transform of $C_{e0}(t)$ evaluated at $s = -i\delta$, where $\delta = \omega - \omega_e$, i.e., 
\begin{equation}\label{eq:app cg1 inf laplace}
  C_{g1}^\infty(\delta) = -i\sqrt{\mathcal{K}(\delta)}
  \;\tilde{C}_{e0}(-i\delta)
  = -i\sqrt{\mathcal{K}(\delta)}\;
  \frac{Q(-i\delta)}{P(-i\delta)}\,,
\end{equation}
using the rational form $\tilde{C}_{e0}(s) = Q(s)/P(s)$ derived in Appendix~\ref{app:laplace solution}.

For a single Lorentzian kernel centered at the emitter frequency ($\Omega_1 = \omega_e$, so $\tilde{B} = B$), the polynomials are $Q(s) = s + B$ and $P(s) = s^2 + Bs + A$. The stationary photon intensity simplifies to
\begin{equation}\label{eq:app photon single L}
  |C_{g1}^\infty(\delta)|^2 
  = \mathcal{K}(\delta)\;
  \frac{\delta^2 + B^2}
  {\delta^4 + (B^2-2A)\,\delta^2 + A^2}\,.
\end{equation}
This spectrum exhibits a local minimum at $\delta = 0$ (a Rabi doublet) when the curvature of the denominator at the origin is negative, which requires $B^2 - 2A < 0$. Recalling that $A = \int d\omega\,\mathcal{K}(\omega)$ is the total spectral weight of the mode and $2B = \Gamma_\mathcal{K}$ is its full width at half maximum, the splitting condition $2A > B^2$ results directly from the strong coupling criterion \cite{Rema2025}
\begin{equation}
  2\int d\omega\,\mathcal{K}(\omega) 
  > \left(\Gamma_\mathcal{K}/2\right)^2\,.
\end{equation}

\section{\label{app:laplace solution}Laplace transform solution of the IDE}

This appendix gives the explicit Laplace-domain solution of the integro-differential equation~(\ref{eq:Ce0}) for a kernel modeled as a sum of $n$ Lorentzians [Eq.~(\ref{eq:kernel lorentzian})].

\subsection{General $n$-Lorentzian kernel}

With the multi-exponential memory kernel from Eq.~(\ref{eq:memory multimode}), the IDE for the excited-state amplitude $y(t)\equiv C_{e0}(t)$ reads
\begin{equation}\label{eq:app ide n}
  \dot{y}(t) = -\int_0^t d\tau\,
  \sum_{j=1}^n A_j\,
  \mathrm{e}^{-\tilde{B}_j\tau}\;
  y(t-\tau)\,,
\end{equation}
where $\tilde{B}_j = B_j + i(\Omega_j - \omega_e)$ and $y(0) = 1$. Taking the Laplace transform and applying the convolution theorem gives
\begin{equation}\label{eq:app laplace eqn}
  s\,Y(s) - 1 
  = -\sum_{j=1}^n \frac{A_j}{s + \tilde{B}_j}\;Y(s)\,,
\end{equation}
which can be solved algebraically:
\begin{equation}\label{eq:app Y rational}
  Y(s) = \frac{Q(s)}{P(s)}\,,
\end{equation}
where the numerator polynomial is
\begin{equation}\label{eq:app Q def}
  Q(s) = \prod_{k=1}^n (s + \tilde{B}_k)\,,
\end{equation}
and the denominator polynomial is
\begin{equation}\label{eq:app P def}
  P(s) = s\,Q(s) + \sum_{k=1}^n A_k\,Q_k(s)\,,
\end{equation}
with $Q_k(s) = \prod_{j\neq k}(s + \tilde{B}_j)$. The polynomial $P(s)$ has degree $n+1$, so the system has $n+1$ poles.

The time-domain solution is obtained by partial-fraction decomposition. Let $\{s_m\}_{m=1}^{n+1}$ denote the roots of $P(s)$. If all roots are simple (no degeneracies), the residue theorem gives~\cite{Polyanin2008handbook}
\begin{equation}\label{eq:app residue solution}
  C_{e0}(t) = \sum_{m=1}^{n+1}
  \frac{Q(s_m)}{P'(s_m)}\;\mathrm{e}^{s_m t}\,,
\end{equation}
where $P'(s) = dP/ds$. This is Eq.~(\ref{eq:residues}) of the main text. Each pole $s_m = -\gamma_m + i\omega_m$ contributes an exponentially damped oscillation at frequency $\omega_m$ with decay rate $\gamma_m$. Complex-conjugate pairs with $|\omega_m| > \gamma_m$ produce damped Rabi oscillations; purely real poles (or pairs with $|\omega_m| \ll \gamma_m$) produce exponential decay.

\subsection{Explicit form for $n=1$}

For a single Lorentzian at $\Omega_1 = \omega_e$ (so $\tilde{B}_1 = B$):
\begin{align}
  Q(s) &= s + B\,,\\
  P(s) &= s^2 + Bs + A\,.
\end{align}
The two roots are $s_\pm = -B/2 \pm \sqrt{B^2/4 - A}$. Defining $b = \sqrt{A - B^2/4}$, the residue formula~(\ref{eq:app residue solution}) yields
\begin{equation}\label{eq:app Ce0 1L}
  C_{e0}(t) = \mathrm{e}^{-Bt/2}
  \left[\cos(bt) + \frac{B}{2b}\sin(bt)\right]\,,
\end{equation}
which is Eq.~(\ref{eq:Ce0 analytical}).

\subsection{Extension to arbitrary $n$}

The construction generalizes directly to arbitrary $n$: the denominator $P(s)$ is always a polynomial of degree $n+1$, and the solution is a sum of $n+1$ exponentials with complex frequencies determined by its roots. For the kernel fits used in this work ($n=7$ for the bare sphere, $n=8$ for the core--shell), the roots are obtained numerically. \notablue{$\mathcal{K}(\omega)$ is the quantity computed from the Mie expansion of the dyadic Green function,
whereas the multi-Lorentzian form is a finite representation of it obtained by fitting. We use it because the Laplace inversion can then be carried out analytically. In all cases considered here the roots satisfy $\mathrm{Re}(s_m)<0$ and $|C_{e0}(t)|^2$ remains bounded by unity, as required.}


\twocolumngrid

\bibliographystyle{apsrev4-1}
\bibliography{macroplex-v2}

\end{document}